\documentclass[aps,pra,twocolumn,groupedaddress]{revtex4-2}
\usepackage[colorlinks=true,linkcolor=blue]{hyperref}
\usepackage{amsmath,amssymb}
\usepackage{bm}

\newcommand{\nn}{\nonumber}
\def\/{\over}
\newcommand{\bra}[1]{\langle#1|}
\newcommand{\ket}[1]{|#1\rangle}
\newcommand{\bea}{\begin{eqnarray}}
\newcommand{\eea}{\end{eqnarray}}
\newcommand{\beq}{\begin{equation}}
\newcommand{\eeq}{\end{equation}}

\begin{document}

%\preprint{}


\title{Resonance interaction due to quantum coherence}

\author{Jiawei Hu}
\email[]{jwhu@hunnu.edu.cn}
\affiliation{Department of Physics, Synergetic Innovation Center for Quantum Effect and Applications, and Institute of Interdisciplinary Studies, Hunan Normal University, Changsha, Hunan 410081, China}
\author{Hongwei Yu}
\email[Corresponding author: ]{hwyu@hunnu.edu.cn}
\affiliation{Department of Physics, Synergetic Innovation Center for Quantum Effect and Applications, and Institute of Interdisciplinary Studies, Hunan Normal University, Changsha, Hunan 410081, China}

\begin{abstract}

The interaction energy between two atoms is crucially dependent on the quantum state of the two-atom system. In this paper, it is demonstrated that a steady resonance interaction energy between two atoms exists when the atoms are in a certain type of coherent superposition of single-excitation states. 
The interaction is tree-level classical in the sense of the Feynman diagrams. A quantity called quantum classicality is defined in the present paper, whose nonzero-ness ensures the existence of this interaction. The dependence of the interatomic interaction on the quantum nature of the state of the two-atom system may potentially be tested with Rydberg atoms.

\end{abstract}

\maketitle

\section{Introduction}

The quantization of electromagnetic fields leads to the prediction of many interesting effects, one of which is the interatomic interaction between neutral atoms \cite{craig1998molecular,salam2009molecular}. Classically, there should be no electromagnetic interaction between neutral non-polar atoms. Quantum mechanically, one atom interacts with the fluctuating vacuum electromagnetic field and a radiative field is induced, which acts on the other atom, and vice versa.  An interatomic interaction is thus generated.

The behaviors of the interatomic interaction energy are significantly different when the atoms are in different quantum states \cite{craig1998molecular,salam2009molecular,passante2018dispersion,andrews2004virtual,cp,McLone1965,Gomberoff1966,Power1993pra,Power1993Chem,Power1995,Rizzuto2004, Sherkunov2007,Preto2013,Donaire2015,Milonni2015,Berman2015,Jentschura2017,Donaire21,Donaire21-2}. For example, when both of the two atoms are in the ground states, the interaction potential is the $r^{-6}$ van der Waals interaction in the near-zone limit  $\omega_0r\ll1$, and the $r^{-7}$ Casimir-Polder interaction in the far-zone limit $\omega_0r\gg1$, where $r$ is the interatomic separation, and $\omega_0$ the energy-level spacing of the atoms~\cite{cp}. When one atom is in its ground state while the other is excited and the excitation transfer is not considered, the interaction behaves as $r^{-6}$ and $r^{-2}$ in the near and far regimes respectively \cite{Power1993pra,Power1993Chem,Power1995}. In either case, the fourth-order perturbation calculation is required, and the result is proportional to the square of the fine structure constant, which can be understood as an exchange of two virtual photons according to the rules of the Feynman diagram. 
The situation is different when the excitation transfer is considered.  Resonance interaction exists when one of a pair of identical atoms is excited, and the excitation is transferred back and forth between the two atoms \cite{craig1998molecular,salam2009molecular,passante2018dispersion,andrews2004virtual}. In many cases of interest, the period of the excitation exchange is shorter than the resolving time, so the existence of a steady resonance interaction energy is of particular interest. In this regard, it is found that when the two atoms are prepared in a symmetric or antisymmetric superposition of the single-excitation states, which is maximally entangled, there exists a steady resonance interaction energy which is proportional to a single power of the fine structure constant, and behaves as $r^{-3}$ and $r^{-1}$ in the near and far regimes respectively \cite{craig1998molecular,salam2009molecular}. 
More importantly, the second-order perturbation suffices for the derivation, and the resonance interaction is proportional to the first power of the fine structure constant, which can be understood as a tree-level classical interaction according to the Feynman rules. 
Since the interatomic interaction is crucially dependent on the quantum state of the two-atom system, a question  arises naturally as to what the criterion is for a steady resonance interaction energy to exist between two atoms in a quantum state. 
Also, if resonance interaction exists for a two-atom system in a state other than the symmetric and antisymmetric states, 
it is not clear how large the interaction is, or in other words,  how the resonance interaction is related to the quantum state of the two-atom system.

In this paper, we investigate the interatomic interaction energy between two identical two-level atoms based on the method proposed by Dalibard, Dupont-Roc, and Cohen-Tannoudji (DDC)~\cite{ddc82,ddc84}, which has recently been generalized from the second order to the fourth order to study the interatomic interaction \cite{Passante14,Menezes17,Zhou21}. In particular, we focus on when the second-order calculation is sufficient to give a nonzero steady resonance interatomic interaction. The Einstein summation convention for repeated indices is applied in the present paper, and the Latin indices run from 1 to 3. Natural units with $\hbar=c=\varepsilon_0=1$ are used in the present paper, where $\hbar$ is the reduced Planck constant, $c$ the speed of light, and $\varepsilon_0$ the vacuum permittivity.

\section{The basic formalism}

The model we consider consists of two identical two-level atoms $A$ and $B$  interacting with the electromagnetic field in vacuum. The Hamiltonian of the two-atom system is
\begin{equation}\label{}
H_S=\omega_0S_z^A+\omega_0S_z^B,
\end{equation}
where $\omega_0$ is energy-level spacing of the atoms, $S_z={1\/2}\left(\ket{e}\bra{e}-\ket{g}\bra{g}\right)$, with $|e\rangle$ and $|g\rangle$ being the excited and the ground states respectively. 
The Hamiltonian of the electromagnetic field can be written as
\begin{equation}
H_F=\sum_{\mathbf{k}\lambda}\omega_{k}a^{\dag}_{\mathbf{k}\lambda}a_{\mathbf{k}\lambda}\,,
\end{equation}
where $\mathbf{k}$ and $\lambda$ denote the wave vector and polarization respectively. 
The dipole interaction between the atoms and the fluctuating electromagnetic field can be described with the following interaction Hamiltonian,
\begin{equation}
H_I=-{\boldsymbol D_{A}}\cdot{\boldsymbol E}(x_A(\tau))-{\boldsymbol D_{B}}\cdot{\boldsymbol E}(x_B(\tau))\,,
\end{equation}
where $\boldsymbol D$ is the electric dipole moment operator of the atoms, and $\boldsymbol{E}(x(\tau))$ is the electric field operator
\begin{equation}
\boldsymbol{E}(x(\tau))=-i\sum_{\bf{k}\lambda}\sqrt{\frac{2\pi k}{V}}\boldsymbol{f}_{\bf{k}\lambda}(\boldsymbol{r})(a_{\bf{k}\lambda}(\tau) - a^{\dagger}_{\bf{k}\lambda}(\tau))\, ,
\end{equation}
with ${\boldsymbol f}_{\bf k\lambda}(\boldsymbol r)$ being the field mode functions, and $V$ the quantization volume. Here $x_A(\tau)$ and $x_B(\tau)$ describes the trajectories of atom $A$ and atom $B$ respectively, and $\tau$ is the proper time.

In the following, we exploit the DDC formalism \cite{ddc82,ddc84} to investigate the resonance interaction between the two atoms, which allows us to separate the contributions of the vacuum fluctuations and radiation reaction to the interaction potential respectively. In the second-order DDC approach, the time evolution of the atomic observables is governed by an effective Hamiltonian, which is found to be \cite{Rizzuto16,Zhou21,zhou2018vacuum}
\begin{equation}
(H^{\text{eff}})_{A}=(H^{\text{eff}}_{A})_{\text{vf}}+(H^{\text{eff}}_{A})_{\text{rr}},
\end{equation}
where
\begin{eqnarray}\label{Hvf}
(H^{\text{eff}}_{A})_{\text{vf}}&=&-{i\/2}\int^{\tau}_{\tau_0}d\tau'C^F_{ij}(x_A(\tau),x_A(\tau'))\nn\\
&&\times[D^{A}_i(\tau),D^{A}_j(\tau')]\,,
\end{eqnarray}
\begin{eqnarray}\label{Hrr}
(H^{\text{eff}}_{A})_{\text{rr}}&=&-{i\/2}\int^{\tau}_{\tau_0}d\tau'\big(\chi^F_{ij}(x_A(\tau),x_A(\tau'))\{D^{A}_i(\tau),D^{A}_j(\tau')\}\nn\\
&&+\chi^F_{ij}(x_A(\tau),x_B(\tau'))\{D^{A}_i(\tau),D^{B}_j(\tau')\}\big)\,,
\end{eqnarray}
are associated with the contributions from  vacuum fluctuations and radiation reaction respectively. 
Here 
\begin{eqnarray}
&&C^F_{ij}(x(\tau),x(\tau'))={1\/2}\langle0|\{E_{i}(x(\tau)),E_{j}(x(\tau'))\}|0\rangle\,,\\
&&\chi^F_{ij}(x(\tau),x(\tau'))={1\/2}\langle0|[E_{i}(x(\tau)),E_{j}(x(\tau'))]|0\rangle\,,
\label{chif}
\end{eqnarray}
are the symmetric and antisymmetric  field correlation functions respectively, with $[..., ... ]$ and $\{..., ...\}$ being the commutator and anticommutator, and $|0\rangle$ the vacuum state of the electromagnetic field. 
The parameter $\tau_0$ in Eqs. \eqref{Hvf} and \eqref{Hrr} is the time when the interaction between the atoms and the field is switched on. Since the environment considered here is stationary in the sense that the correlation functions of the field are invariant under temporal translations, this parameter does not play any role in the calculation. 
Similarly, the effective Hamiltonian   for atom $B$ can be derived by exchanging the subscript $A$ with $B$ in the equations above.

We assume that the quantum state of the two-atom system can be described with a density matrix $\rho_{AB}$. Taking the expectation values of Eqs. \eqref{Hvf} and \eqref{Hrr} with respect to the atomic state $\rho_{AB}$, the  energy shift of the atomic system due to vacuum fluctuations and  radiation reaction can be obtained respectively as, 
\begin{eqnarray}\label{Evf}
&&(\delta E_A)_{\text{vf}}=-i \int^{\tau}_{\tau_0}d\tau'C^F_{ij}(x_A(\tau),x_A(\tau'))\chi^{AA}_{ij}(\tau,\tau')\,,\quad\\
&&(\delta E_A)_{\text{rr}}=-i\int^{\tau}_{\tau_0}d\tau'
\big[\chi^F_{ij}(x_A(\tau),x_A(\tau'))C^{AA}_{ij}(\tau,\tau')\nn\\
&&\qquad\qquad\quad+\chi^F_{ij}(x_A(\tau),x_B(\tau'))C^{AB}_{ij}(\tau,\tau')\big]\,,
\label{Err}
\end{eqnarray}
where 
\begin{eqnarray}\label{Cij}
&&C^{AB}_{ij}(\tau,\tau')={1\/2}\,{\rm Tr}\left(\rho_{AB}\, \{D^{A}_{i}(\tau),D^{B}_{j}(\tau')\}\right)\,,\\
&&\chi^{AB}_{ij}(\tau,\tau')={1\/2}\,{\rm Tr}\left(\rho_{AB}\, [\,D^{A}_{i}(\tau),D^{B}_{j}(\tau')\,]\right)\,,
\end{eqnarray}
are the symmetric and antisymmetric statistical functions of the atoms respectively. From Eqs. \eqref{Evf} and \eqref{Err}, it is clear that the only term relevant to the interatomic interaction is the second term in Eq. \eqref{Err}, since the other terms do not depend on the interatomic separation.
Let us note that the fact that only the radiation reaction term
contributes to the resonance interaction between two correlated atoms was first shown in Refs. \cite{Rizzuto16,zhou2018vacuum}. 
Adding up the contributions from atom $A$ and $B$, the total  interatomic interaction energy can be obtained as
\begin{eqnarray}\label{Etot}
\delta E&=&-i\int^{\tau}_{\tau_0}d\tau'\chi^F_{ij}(x_A(\tau),x_B(\tau'))C^{AB}_{ij}(\tau,\tau')\nn\\&&+(A\rightleftharpoons B)\,.
\end{eqnarray}

\section{The second-order resonance interaction}

In the previous section, we have derived an interatomic interaction based on the second-order DDC approach, as shown in Eq. \eqref{Etot}. However, this result is only formal in the sense that it may actually be zero. An obvious example is when the two atoms are in their ground states, the interaction is the fourth-order Casimir-Polder interaction~\cite{cp}. In the following, we investigate the criterion for a nonzero second-order resonance interaction. 

In order to obtain the criterion for the existence of the tree-level resonance interaction, we start from the most generic quantum state of a  two-atom system without any predetermined assumptions, 
which can be written in the basis $\{ |ee\rangle, |eg\rangle, |ge\rangle, |gg\rangle \}$ as 
\begin{equation}\label{rho}
\rho_{AB}=
\left(
\begin{array}{cccc}
\rho_{11} & \rho_{12} & \rho_{13} & \rho_{14} \\
\rho_{21} & \rho_{22} & \rho_{23} & \rho_{24} \\
\rho_{31} & \rho_{32} & \rho_{33} & \rho_{34} \\
\rho_{41} & \rho_{42} & \rho_{43} & \rho_{44} 
\end{array}\right).
\end{equation}
Taking Eq. \eqref{rho} into Eq. \eqref{Cij}, it is straightforward to obtain
\begin{eqnarray}\label{Cij-2}
C^{AB}_{ij}(\tau,\tau')&=&
\rho_{23}\,d_i^{A}d_j^{B*}e^{-i\omega_0\Delta\tau}+\rho_{32}\,d_i^{A*}d_j^{B}e^{i\omega_0\Delta\tau}\nn\\
&&+\rho_{14}\,d_i^{A}d_j^{B}e^{i\omega_0\Delta\tau-2i\omega_0\tau}\nn\\
&&+\rho_{41}\,d_i^{A*}d_j^{B*}e^{-i\omega_0\Delta\tau+2i\omega_0\tau},
\end{eqnarray}
where $d_i=\langle g|D_i|e\rangle$, and  $\Delta\tau=\tau-\tau'$. 
Here, the correlation function of the atoms Eq.~\eqref{Cij-2} contains  terms depending on $\tau+\tau'$, which indicates that it is not homogeneous in time when $\rho_{14}$ and $\rho_{41}$ are nonzero, i.e., there exists a superposition of $|gg\rangle$ and $|ee\rangle$. 
The reason for this is that, when $\rho_{14}$ and $\rho_{41}$ are nonzero, the state of the two-atom system $\rho_{AB}$ does not commute with the Hamiltonian of the two-atom system $H_S$, i.e., $[\rho_{AB},H_S]\neq0$. According to the Heisenberg equation $\dot{\rho}_{AB}=-i[\rho_{AB},H_S]$,  the quantum state $\rho_{AB}$ evolves even if  the two-atom system were isolated from the fluctuating electromagnetic fields. This is why the process is not stationary when $\rho_{14}$ and $\rho_{41}$ are nonzero. 
Consequently,  the  interatomic interaction energy is time-dependent. Nevertheless, as will be discussed later, these oscillating terms do not contribute to the steady resonance interaction. 
Also, the correlation function Eq.~\eqref{Cij-2} is independent of $\rho_{11}$ and $\rho_{44}$, which indicates that the  second-order  resonance interaction does not exist when both of the two atoms are in their ground or excited states.

The Wightman function of the electric field reads (See, e.g., Ref. \citep{Greiner}),
\begin{widetext}
\begin{eqnarray}
\langle0|{E}_{i}(x(\tau)){E}_{j}(x'(\tau'))\vert0\rangle=-{1\/4\pi^2}(\delta_{ij}\partial_0\partial_{0'}-\partial_i\partial_{j'})
{1\/{(\tau-\tau'-i\epsilon)^2-(x-x')^2-(y-y')^2-(z-z')^2}}\,,
\end{eqnarray}
where $\epsilon\to0^+$. We assume that the atoms are  a distance $r$ apart along the $z$-axis, so the trajectories of the two atoms can be written as
\begin{eqnarray}
&&\tau_A=\tau,\;x_A=0,\;y_A=0,\;z_A=0,\\
&&\tau_B=\tau',\;x_B=0,\;y_B=0,\;z_B=r.
\end{eqnarray}
Then, the antisymmetric field correlation function \eqref{chif} is found to be \cite{Rizzuto16,zhou2018vacuum},
\begin{eqnarray}
\label{chif-2}
&&\chi^{F}_{ij}(x_A(\tau),x_B(\tau'))=\int^{\infty}_{0}\frac{d\omega}{8\pi^2}
\biggl\{\biggl[(\delta_{ij}-3n_in_j)\biggl({\sin\omega r\/r^3}-{{\omega}\cos\omega r\/r^2}\biggr)-(\delta_{ij}-n_in_j){\omega^2\sin\omega r\/r}\biggr]\biggr\}(e^{i\omega\Delta\tau}-e^{-i\omega\Delta\tau}).\quad~
\end{eqnarray}
\end{widetext}

Plugging Eqs. (\ref{Cij-2})  and (\ref{chif-2}) into Eq. (\ref{Etot}), with a substitution $\Delta\tau=\tau-\tau'$ and an extension of the range of integration to infinity, the interatomic interaction between two atoms can be obtained. The first two terms in Eq. (\ref{Cij-2}) contribute a  time-independent interatomic interaction energy, 
\begin{eqnarray}\label{Etot-2}
\delta E=\big(\rho_{23}\,d_i^{A}d_j^{B*} + \rho_{32}\,d_i^{A*}d_j^{B}\big){\cal V}_{ij},
\end{eqnarray}
where 
\begin{eqnarray}\label{Vij}
{\cal V}_{ij}&=&\frac{1}{4\pi r^3}
       \big[ (\delta_{ij}-3n_i n_j)(\cos \omega_0 r + \omega_0 r\sin \omega_0 r)\nn\\
      &&\qquad\qquad- (\delta_{ij}-n_i n_j) \omega_0^2 r^2 \cos \omega_0 r\big],
\end{eqnarray}
is the dipole-dipole interaction tensor (See, e.g., Refs. \cite{craig1998molecular,salam2009molecular}), 
while the last two terms lead to a time-dependent one 
$\delta E (\tau)=\big(\rho_{41}\,d_i^{A*}d_j^{B*}e^{2i\omega_0\tau} + \rho_{14}\,d_i^{A}d_j^{B}e^{-2i\omega_0\tau}\big){\cal V}_{ij}$, 
which rapidly oscillates between attractive and repulsive. In many occasions, the time characterized by the inverse of the transition frequency is shorter than the resolving time, so the oscillating terms do not contribute to the steady resonance interaction energy in the sense that the time average $\frac{1}{T}\int_0^T d\tau\,\delta E (\tau)$ is zero. 
We observe from Eq. \eqref{Etot-2} that the second-order resonance interaction is crucially related to the density matrix element $\rho_{23}$, which is associated with the quantum coherence between the single-excitation states $\ket{ge}$ and $\ket{eg}$. 
Furthermore, using the property ${\cal V}_{ij}={\cal V}_{ji}$, and recalling that the two atoms are assumed to be identical such that $d_i^{A}=d_i^{B}$, it can be shown that  only the real part of $\rho_{23}$ is relevant to the second-order interatomic interaction. This prompts us  to define a new physical quantity
\begin{equation}
{\cal Q}\equiv {\rm Re}\, (\rho_{23})
\end{equation}
We call ${\cal Q}$  {\it quantum classicality} to characterize the quantum coherence that induces the tree-level classical resonance interaction. The  interaction energy between the two atoms can then be reformulated as
\begin{equation}\label{Etot-3}
\delta E=2{\cal Q}\,{\rm Re}(d_i^{A}d_j^{B*})\, {\cal V}_{ij}.
\end{equation}
In Ref. \cite{l1norm}, it is proposed to measure the quantum coherence of a quantum system with the sum of the modules of all the off-diagonal elements, i.e., ${\cal C}_{l_1}=\sum_{i\neq j}|\rho_{ij}|$, which is known as $l_1$-norm. Obviously, a nonzero quantum coherence as quantified by $l_1$-norm   is a necessary but not sufficient condition for the existence of a second-order resonance interaction.

In the following, we show some  explicit examples after the general derivation. 
 
{\it 1. Pure-state example.}---Consider the following superposition state, 
\begin{equation}\label{ent}
|\psi\rangle=\sin\theta\,|ge\rangle +\cos\theta\, e^{i\phi} |eg\rangle,
\end{equation}
where $\theta,\phi\in (-\pi,\pi]$ are the weight parameter and the phase parameter respectively. Here it is easy to find that quantum classicality ${\cal Q}=\frac{1}{2}\sin2\theta\,\cos\phi$, and the resonance interaction energy is 
\begin{eqnarray}\label{int-ent}
\delta E=\sin2\theta\cos\phi\, {\rm Re}(d_i^{A}d_j^{B*})\, {\cal V}_{ij}.
\end{eqnarray}
It is straightforward to examine that  the quantum coherence  of the two-atom system in the pure state \eqref{ent} measured by $l_1$-norm ${\cal C}_{l_1}$ \cite{l1norm} and the quantum entanglement measured by concurrence ${\cal C}$ \cite{concurrence} are equal, i.e., ${\cal C}={\cal C}_{l_1}=|\sin2\theta\,|$. As long as $\phi\neq\pm\pi/2~ (\cos\phi\neq0)$, there exists a resonance interaction between the two atoms,  the  magnitude of which is proportional to the amount of quantum coherence  measured by $l_1$-norm and  quantum entanglement measured by concurrence in this case. That is, when the two-atom system is prepared in the superposition state \eqref{ent}, quantum entanglement is a necessary condition for a nonzero resonance interaction. When $\phi=0$ and $\theta=\pm\pi/4$, the result reduces to the well-known resonance interaction for symmetric and antisymmetric states \cite{craig1998molecular,salam2009molecular}.

{\it 2. Mixed-state example.}---Consider the Werner state \cite{Werner} of the following form, 
\begin{equation}\label{werner}
\rho=\frac{1-p}{4}\,\mathbb{I}+p|\Psi_+\rangle\langle\Psi_+|,
\end{equation}
with $\mathbb{I}$ being the $4\times4$ unit matrix, $|\Psi_+\rangle=\frac{1}{\sqrt{2}}(|ge\rangle + |eg\rangle)$ the symmetric state, and $p$ a positive parameter. 
Here, ${\cal Q}=p/2$, and the resonance interaction energy is 
\begin{equation}\label{werner2}
\delta E=p\,{\rm Re}(d_i^{A}d_j^{B*})\,{\cal V}_{ij}.
\end{equation}
In this case, the concurrence is ${\cal C}=\max\{\frac{3p-1}{2},0\}$, and the $l_1$-norm is ${\cal C}_{l_1}=p$. When $0<p\leq\frac{1}{3}$, two atoms are not entangled, while there exists quantum coherence between the single-excitation states $\ket{ge}$ and $\ket{eg}$ as measured by $l_1$-norm   and a nonzero quantum classicality  ${\cal Q}$. 
In contrast to the pure-state case, entanglement is not necessary for the resonance interaction here.

\section{Discussions}

A few comments are now in order. 

First, for the question we raised at the beginning, it is now clear that quantum entanglement is neither a necessary nor a sufficient condition for a nonzero second-order resonance interaction. Also, quantum coherence, e.g., as measured by $l_1$-norm, is only  necessary but not sufficient. In the present paper, we show that the  necessary and sufficient condition for the existence of a second-order resonance interaction between two identical atoms is when the quantum classicality defined as the real part of the density matrix element $\rho_{23}$  is nonzero. When the two-atom system is  in a state $|\Psi\rangle$ such that $\rho_{23}$  is pure imaginary, e.g., $|\Psi\rangle=\frac{1}{\sqrt{2}}(|ge\rangle+i|eg\rangle)$, $C^{AB}_{ij}(\tau,\tau')$ is nonzero. However, since $C^{BA}_{ij}(\tau,\tau')$ is the complex conjugate of $C^{AB}_{ij}(\tau,\tau')$ according to its definition \eqref{Cij}, contributions from the two atoms cancel out, and   the second-order interatomic interaction of the two-atom system is zero.

Second, the tensor ${\cal V}_{ij}$ in Eq. \eqref{Vij} is the dipole-dipole interaction tensor describing the classical dipole-dipole interaction, which  behaves as $r^{-3}$ in the near region ($\omega_0 r\ll1$), and $r^{-1}$ in the far region ($\omega_0 r\gg1$). 
In many occasions, $\omega_0$ tends to be very small, so the resonance interaction remains  nonretarded over all experimentally relevant distance ranges where the interaction appreciably differs from zero. 
Furthermore, the resonance interaction \eqref{Etot-2} is proportional to the first power of the fine structure constant $\alpha$, which can be regarded as a result of the exchange of a single photon according to the Feynman rules. 
Therefore, it is a tree-level classical interaction in the sense of Feynman diagrams.

Third, the resonance interaction, although tree-level classical in the sense of the Feynman diagrams,  is intimately related to the quantum nature of the atoms. 
In particular, here we assume that the atoms are non-polar ones, in the sense that the quantum expectation values of the dipole moments of the atoms ${\rm Tr}(\rho_{A}D^{A}_{i})={\rm Tr}(\rho_{B}D^{B}_{i})=0$, where $\rho_{A(B)}=\text{Tr}_{B(A)}\rho_{AB}$ is the reduced density matrix of atom $A(B)$. Note that, from Eqs. \eqref{ent} and \eqref{werner}, it is easy to check that the atoms are non-polar in our examples. Classically, there should be no electromagnetic interactions between neutral nonpolar objects. So we call the interaction here a tree-level classical resonance interaction brought by quantum coherence, and name the quantity that ensures  the existence of this interaction when it is nonzero as quantum classicality. In other words, the criterion is that the quantum classicality must be nonzero for such  an interaction to exist.

Finally, our result shows that the resonance interaction between two atoms can be controlled by the quantum nature of the two-atom state, which may potentially be verified with the Rydberg atoms. Rydberg atoms are atoms with a large principal quantum number $n$, and have strong dipole-dipole interaction scaling as $n^4$ \cite{Gallagher_1988,Saffman2010}. 
For example, for Rb atoms with principal quantum number $n\sim 60$ separated at several microns, the resonance interaction $\delta E/\hbar$ is of the order of $\sim 10$ MHz \cite{Saffman2010}.
To detect the state-dependent resonant interaction, two atoms are first trapped in microscopic optical traps separated by $r$, and then optically pumped into the Rydberg state. For a detailed description of a typical experimental setup concerning Rydberg atoms please refer to Ref. \cite{Browaeys13}. Then, starting from the state $|eg\rangle$, one can create any coherent superposition of the single-excitation states $|ge\rangle$ and $|eg\rangle$ by applying a state-selective light shift using an addressing beam as demonstrated in a recent work \cite{Sylvain17}. 
The exaggerated interaction and manipulable quantum state make us believe that it is hopeful to test the dependence of the resonance interaction on the nature of the two-atom quantum state with Rydberg atoms.

\section{Summary}

In this paper, we have shown that a steady resonance interaction between two neutral nonpolar atoms can be induced when the atoms are in a state with nonzero quantum classicality, a quantity defined in the present paper to characterize this property. 
The interaction is tree-level classical in the sense of the Feynman diagrams. The dependence of the resonance interaction on the quantum nature of the two-atom state can hopefully be tested with Rydberg atoms.

%\appendix
%\section{}

\begin{acknowledgments}
The authors would like to thank Yongshun Hu for helpful discussions. 
This work was supported in part by the NSFC under Grants No. 11805063, No. 11690034, and No. 12075084. 
\end{acknowledgments}

\bibliography{ref}

\end{document}